\documentclass{appolb}
\usepackage{graphicx}

\usepackage{graphicx}%
\usepackage{multirow}%
\usepackage{amsmath,amssymb,amsfonts}%
\usepackage{amsthm}%
\usepackage{mathrsfs}%
\usepackage[title]{appendix}%
\usepackage{xcolor}%
\usepackage{textcomp}%
\usepackage{manyfoot}%
\usepackage{booktabs}%
\usepackage{algorithm}%
\usepackage{algorithmicx}%
\usepackage{algpseudocode}%
\usepackage{listings}%

\definecolor{orange}{RGB}{255, 121, 0}
\definecolor{dgreen}{rgb}{0.0,0.7,0.0}

\RequirePackage{fix-cm}

\newcommand{\bsub}{\begin{subequations}}
\newcommand{\esub}{\end{subequations}}

\newcommand{\beq}{\begin{equation}}
\newcommand{\eeq}{\end{equation}}
\newcommand{\beqn}{\begin{equation*}}
\newcommand{\eeqn}{\end{equation*}}

\newcommand{\bal}{\begin{align}}
\newcommand{\eal}{\end{align}}
\newcommand{\baln}{\begin{align*}}
\newcommand{\ealn}{\end{align*}}

\newcommand{\bea}{\begin{eqnarray}}
\newcommand{\eea}{\end{eqnarray}}
\newcommand{\bean}{\begin{eqnarray*}}
\newcommand{\eean}{\end{eqnarray*}}

\newcommand{\bga}{\begin{gather}}
\newcommand{\ega}{\end{gather}}
\newcommand{\bgan}{\begin{gather*}}
\newcommand{\egan}{\end{gather*}}

\newcommand{\bmat}{\begin{pmatrix}}
\newcommand{\emat}{\end{pmatrix}}

\def\e{\varepsilon}

\usepackage{graphicx}
\usepackage{enumitem}
\usepackage{amsmath}
\usepackage{hyperref}
 \hypersetup{
     colorlinks=true,
     linkcolor=blue,
     filecolor=blue,
     citecolor = black,      
     urlcolor=blue,
     }
 \usepackage{mathptmx}     
 \usepackage{tikz}
 \usetikzlibrary{shapes.geometric, arrows}

\begin{document}
\title{{Inflation and Dark Energy from a Covariant Elastic Medium}}
\author{Mathieu Beau
\address{Department of Physics, University of Massachusetts, Boston, MA, USA}
}
\maketitle
\begin{abstract}
{In this article, we propose a unified framework for cosmological expansion and
inflation at the level of background dynamics, by modeling both the inflaton field and dark energy as a
four-dimensional continuous medium, whose elastic deformation is described by a
covariant vector field. Focusing on homogeneous and isotropic background
cosmology, we show that for a bulk modulus
$K = 1.64 \times 10^{109}~\text{N}\cdot\text{m}^{-2}$, the dark energy density
decreases by a factor of $\sim 10^{122}$ while the scale factor expands
$10^{28}$ times over $\sim 10^{-42}$ seconds during primordial inflation.
For illustrative parameter values, our analysis suggests three potential new
physical phenomena for future investigation, including longitudinal elastic
modes, frequency redshifts in early-universe light, and improved fits to
supernova curves. At the end of the paper, we discuss the challenges of applying the framework to
inflationary perturbations, particularly the need for a consistent theory
capable of producing a nearly scale-invariant power spectrum, as well as of
addressing reheating, and identify these as key directions for future work.}

\end{abstract}

\section{Introduction}

Recent observations of gravitational waves by LIGO Scientific Collaboration and Virgo Collaboration have opened new frontiers in the search for invisible forms of energy and phenomena that were beyond the current observable universe using optical instruments \cite{LIGO16,LIGO16(2),LIGO17}. In addition, modern cosmology has a plethora of precise observations, such as supernovae luminosity distances \cite{Riess98} and Cosmic Microwave Background (CMB) \cite{COBE92,WMAP,Planck13}. These results have been used to refine the prediction of the $\Lambda$CDM model parameters \cite{Hubble16,Planck16}, even though there is a significant discrepancy in the value of the Hubble constant between the results from Planck mission, the Hubble Space Telescope \cite{Hubble16,Hubble18}, and most recently the Dark Energy Spectroscopic Instrument (DESI) collaboration \cite{DESI_APS24,DESI24}. The LIGO Scientific Collaboration and the Virgo Collaboration have also predicted a value of the Hubble constant \cite{LIGOHubble17,LIGOHubble21,LIGOHubble21corr} and, most certainly, future data obtained using gravitational detectors (aLIGO, eLISA, Einstein Telescope) will give new horizon for cosmology \cite{Sesana16,Askar19,Amaro23}. 
However, important theoretical questions remain unsolved and must be addressed to clarify the interpretation and understanding of future observations. Consider, for instance, the puzzling cosmological constant problem \cite{Weinberg89}, i.e., the $120$ orders of magnitude discrepancy between the theoretical prediction and the observed value of the cosmological constant. 

Different theoretical investigations are using various approaches to address this issue. The Eckart-Israel formalism \textit{viscous fluid model} \cite{Eckart,Israel} was used to develop \textit{dissipative cosmology} \cite{Maartens} where dark energy is described through the concept of \textit{bulk viscosity} \cite{Fabris}. Alternative gravity theories were also developed extensively, e.g., \textit{quintessence} theory \cite{QuintessencePeebles,QuintessenceCaldwell,QuintessenceBrax,QuintessenceReview}, \textit{k-essence} theory \cite{kessence}, scalar field theories with \textit{higher order derivative Lagrangian density} \cite{Anisimov}, \textit{f(R) theories}, and \textit{scalar-tensor theory}, see \cite{AltGravReview} for a comprehensive review of the subject. The possibility of the violation of energy conservation relying on an effective time-dependent cosmological constant 
has been explored \cite{Perez17,Perez18} and also drawn a connection with quantum gravity. An alternative model involving a \textit{negative mass fluid} has been employed to simulate galaxy rotations and cosmological expansion \cite{Farnes}. Another approach that consists of assigning physical property to the space-time continuum, interpreted as an elastic medium \cite{Tartaglia,Millette} or as a solid material \cite{Pearson,Tenev18,David20}. In \cite{Beau15} and \cite{Beau20}, the authors introduced a vector deformation field into the Einstein equations, coupling it with matter-energy through a higher derivative Lagrangian theory.
However, cosmological inflation and expansion mechanisms have not yet been investigated using the approaches mentioned above.
{Unified descriptions of inflation and late-time acceleration have previously been
explored within the scalar-field paradigm, notably in the context of quintessential
inflation (see, e.g., Dimopoulos \cite{Dimopoulos2021}). In such models, a single
scalar degree of freedom drives both the primordial inflationary phase and the
late-time accelerated expansion. In contrast, the present work proposes a
fundamentally different mechanism in which inflation and dark energy arise from
elastic stress--energy associated with the deformation of a four-dimensional
cosmological medium, without invoking scalar fields or scalar potentials.}

In this article, we propose a unifying framework that models both the inflaton field and dark energy within a four-dimensional continuous-medium deformation theory, and we examine the mechanisms governing inflation and cosmological expansion through this approach.
In the standard $\Lambda$CDM picture, the inflaton is the scalar field driving primordial inflation, while dark energy is the agent of late-time acceleration, usually treated as a cosmological constant.
By contrast, in the present elastic-medium framework, both epochs emerge from the same underlying continuum, albeit with vastly different deformation rates.

The content of this article is twofold: we first introduce a model to describe the deformation of a four-dimensional elastic continuous medium that models both the inflaton field and the dark energy, and study the mechanism of both inflation and cosmological expansion within our unifying framework.
We first introduce the general features of the theory and then modify the Friedmann equations obtained for the Friedmann-Lema\^{i}tre-Robertson-Walker (FLRW) metric. We show that the dark energy density is proportional to the product of the bulk modulus $B=-K<0$ and the time-dependent deformation rate $\e(t)$. We study the effects of these two parameters on the luminosity curves of celestial bodies obtained for the dark energy density ratios found by Planck Collaboration and by the Hubble Space Telescope \cite{Planck16,Hubble16}.  We show that for $K\approx 1.64 \times 10^{109}\text{N}\cdot \text{m}^{-2}$ the dark energy density drops by a factor $\sim 10^{122}$, while the scaling factor expands the scale factor expands $10^{28}$ times during a time of the order of $10^{-42}\text{s}$. At the end of the article, we discuss a microscopic interpretation of the inflation mechanism and future investigations.


\section{Dark energy as a cosmological medium - paradigm and theory.}\label{Section:Theory} 

\subsection{Postulates.}

We first propose generalizing to any deformations, i.e., elastic, inelastic, solid, etc. Let us reformulate the hypothesis through these three postulates: \\
\begin{enumerate}[label=\textbf{(P\arabic*)}]
\item Dark energy is considered a continuous medium with specific physical properties, including elasticity, rigidity, stiffness, solidity, shear strength, and incompressibility, characterized by physical constants such as Young's modulus, stiffness, shear modulus, and other relevant parameters.\\

\item The stress resulting from the deformations of this medium is mathematically described by a covariant tensor that alters the internal energy of the system. Consequently, the stress tensor corresponding to the medium is incorporated into the Einstein equations to account for the cost of energy associated with the deformation of the medium.\\

\item The strain field of dark energy generates acceleration currents of the matter-energy contained within the space-time. In other words, the strain field distorts the geodesic paths of all types of matter-energy fields, including massive particles and electromagnetic fields.\\
\end{enumerate}

The first postulate (P1) postulates that dark energy can be described by a four-dimensional continuous elastic theory, which extends beyond the traditional approach of treating dark energy as a constant field in the Einstein equation (despite the various alternative approaches outlined in the introduction). The second postulate (P2) introduces an additional stress-energy tensor (the stress energy caused by the elastic deformation of the medium) into the Einstein equation.

The novelty of our approach lies in the combination of postulates (P2) and (P3), with particular emphasis on postulate (P3), which is less intuitive compared to the second postulate. According to (P2), matter-energy induces a deformation of the continuous medium, while the reciprocal, stated by postulate (P3) suggests that the deformation of the continuous medium influences the trajectories of matter-energy within spacetime.

We interpret postulate (P3) as analogous to Newton's third law of action and reaction in mechanics or Faraday's induction principle, where a change in the deformation field (strain field) leads to an acceleration (in a covariant sense). This analogy underscores the reciprocal actions between matter-energy (electric current) and the strain field (electromagnetic field). 

Mathematically, we demonstrate that the standard General Relativity (GR) framework, where dark energy is modeled as a constant, can be derived as a mathematical limit of our extended GR theory, see Figure \eqref{Fig:ExtendedGR}. We use the term ``extended" to denote that our theory not only introduces additional terms in the stress-energy tensor of the Einstein equation but also incorporates the action of the strain field on matter-energy as per postulate (P3).

In the subsequent subsection, we will present the mathematical expressions corresponding to postulates (P1)-(P3) outlined above.

\begin{figure}
\begin{tikzpicture}[scale=0.7]
    \filldraw[fill=gray!7] (0,-2) ellipse(8.7cm and 4.5cm);
    \filldraw[fill=gray!2,rotate=17, dashed
    ] (-2,-0.75) ellipse(5.8cm and 2.5cm);
    \node (Space) [ellipse, draw,
    minimum width = 1cm, 
    minimum height = 1.5cm, fill=white] at (0,0) {Space-time metric};
    \node (Matter) [ellipse, draw,
    minimum width = 2cm, 
    minimum height = 1.5cm, fill=white] at (-4.5,-3) {Matter-energy};
    \node (Stress) [ellipse, draw,
    minimum width = 1.2cm, 
    minimum height = 1.2cm, fill=white] at (4.9,-3) {Continuous medium};
    \node [text width=2cm,rotate=0] at (-3.3,-0.5) {\Large   GR};
    \node [text width=3cm,rotate=0] at (2,-5) {\Large  Extended GR};
    \draw [latex-latex] (Space) -- node [text width=2.5cm,midway,above,align=center,rotate=35 ] {\small Einstein}
    node [text width=2.5cm,midway,below,align=center,rotate=35 ] {\small Theory}(Matter);
    \draw [latex-latex] (Space) -- node [text width=2.5cm,midway,above,align=center,rotate=-37 ] {\small (P1)} node [text width=2.5cm,midway,below,align=center,rotate=-37 ] {\small (P2)} (Stress);
    \draw [latex-latex] (Stress) -- node [text width=2.5cm,midway,above,align=center,rotate=0 ] {\small (P3)} (Matter);
\end{tikzpicture}
    \caption{\textbf{Diagram representing the relation between the physical concepts (metric, matter-energy, strain/stress) in the general relativity theory (GR) and the extended theory (Extended GR).} In this diagram, we show the interplay of the physical concepts of our theory. The relation between the space-time continuum and matter-energy is described by GR while the effects of the physical properties of the cosmological medium as well as the relation with the metric and the matter-energy are encoded in the three postulates (P1)-(P3). }\label{Fig:ExtendedGR}
\end{figure}

\subsection{Mathematical equations of the model} 

\textit{Strain and stress tensors.} The first postulate (P1) means that the space-time continuum is a physical substance that can be deformed by stretch, compression, and shear. We assume a linear relation between the stress-energy tensor and the strain tensor $\varepsilon_{\mu\nu}$ as 
\beq\label{Section:Theory:Eq:StressTensor1}  
\sigma_{\mu\nu}=C_{\mu\nu\alpha\beta}\ \varepsilon^{\alpha\beta}\ ,
\eeq
where $C_{\mu\nu\alpha\beta}$ is the elasticity tensor that characterizes the physical properties of the 3+1-dimensional continuous medium. Of course, we could extend our model to a non-linear theory, but we leave this for future research work.

By analogy with three-dimensional continuous elastic theory \cite{Beau15,Beau20}, the strain tensor is the symmetrical gradient of the 4-vector deformation $G_\mu$ 
\begin{equation}\label{Eq:StrainTensor}
    \varepsilon_{\mu\nu} = D_\mu G_\nu + D_\nu G_\mu ,
\end{equation}
where $D_\mu$ is the covariant derivative.

\textit{Modified Einstein equations.} Mathematically, the second postulate (P2) states that there is an additional term in the action that characterizes the stress-energy:
\begin{equation}\label{Eq:Action:Modified}
    S = \int \left( R + \sigma + T \right)\sqrt{-g}\ d^4x \ ,
\end{equation}
where $T = g_{\mu\nu} T^{\mu\nu}$ and $\sigma = g_{\mu\nu} \sigma^{\mu\nu}$. 
It comes that the energy corresponding to the elastic deformation of the cosmological continuous medium is encoded in a stress-energy tensor $\sigma_{\mu\nu}$, which is added to Einstein's equations 
\begin{equation}\label{Section:Theory:Eq:ModEinstein} 
R_{\mu}^{\nu}=\frac{8\pi G}{c^4}\left(T_{\mu}^{\nu}+\sigma_{\mu}^{\nu}-\frac{1}{2}\delta_{\mu}^{\nu}\left(T+\sigma\right)\right)\ ,
\end{equation}
where $T_{\mu}^{\nu}=g^{\nu\lambda}T_{\mu\lambda}$ is the momentum-energy tensor. We denote $T=T_{\mu}^\mu$ and $\sigma=\sigma_\mu^\mu$ the traces of the tensors. The meaning of this additional term is that the elastic deformations of the 3+1D medium cost some energy. Another consequence of this modification is that the matter-energy density is generally not conserved:
\begin{equation}\label{Section:Theory:Eq:Divergence} 
D_{\nu}T_{\mu}^{\nu}=-D_{\nu}\sigma_{\mu}^{\nu}\ .
\end{equation}

\textit{Universal coupling between the deformations and the matter-energy within the space-time.} Postulate (P3) indicates that there is a coupling between the deformation vector $G_\mu$ and the covariant acceleration of the matter-energy, i.e., the covariant derivative of matter-energy tensor $D_\mu T^{\mu\nu}$. We posit that this coupling is linear and has the form \cite{Beau20}
\begin{equation}\label{Eq:LangrangianInteractionDefMatter}
    \Lambda_{\text{int}} = -G_{\mu}D_{\nu}T^{\mu\nu}\ , 
\end{equation}
where $\Lambda_{\text{int}}$ is the Lagrangian density, $D_{\nu}T^{\mu\nu}$ is the \textit{4-vector acceleration current} \cite{Beau20}, and $T^{\mu\nu} = \sum_{j=1}^{N} T_{(j)}^{\mu\nu}$ is the sum of the matter-energy tensors. This means that the coupling with the 4-vector deformation is \textit{universal}, in the sense that any type of matter-energy experiences a deflection. From equation \eqref{Eq:LangrangianInteractionDefMatter}, we find that the equations of motion for the $j$th tensor energy are modified as follows
\begin{equation}\label{Eq:DefGeod:General}
    D_{\mu}T_{(j)\ \nu}^{\ \mu}= -\varepsilon_{\mu}^{\gamma}\partial_{\nu}T_{(j)\ \gamma}^{\ \nu}+\Delta^{\mu}_{\mu\gamma}T_{(j)\ \nu}^{\ \gamma}-\Delta^{\gamma}_{\nu\mu}T_{(j)\ \gamma}^{\ \mu}\ ,
\end{equation}
with $\Delta^\gamma_{\mu\nu}\equiv \frac{1}{2}g^{\gamma\lambda}\left(\partial_{\mu}\varepsilon_{\lambda\nu}+\partial_{\nu}\varepsilon_{\lambda\mu}-\partial_{\lambda}\varepsilon_{\mu\nu}\right)$. For example, for a massive particle, the modified geodesic reads $\ddot{x}^\mu +\Gamma^\mu_{\nu\lambda}\dot{x}^\nu\dot{x}^\lambda=-\varepsilon_{\nu}^{\mu}\ddot{x}^\nu -\Delta^\mu_{\nu\lambda}\dot{x}^\nu\dot{x}^\lambda$, see \cite{Beau15,Beau20}. We would like to point out that the deformation of geodesic is formally analog to the linear transformations 
\begin{subequations}
    \begin{equation}\label{Eq:GeoDefx}
        x^\mu \mapsto x'^{\mu}(x)=x^\mu + G^\mu(x) ,
    \end{equation}
    \begin{equation}\label{Eq:GeoDefg}
        g_{\mu\nu}\mapsto g_{\mu\nu}+\varepsilon_{\mu\nu} ,
    \end{equation}
\end{subequations}
where $\varepsilon_{\mu\nu}(x)=D_{\mu}G_{\nu}(x) + D_{\nu}G_{\mu}(x)$. However, it is important to understand that these transformations are not equivalent to the procedure detailed above that leads to the system of equations \eqref{Section:Theory:Eq:Divergence}-\eqref{Eq:DefGeod:General}. Indeed, it is well-known that the standard GR is invariant by diffeomorphism transformations. Using the transformations above one can formally find equation \eqref{Eq:DefGeod:General}, however the physical meaning of the terms in the equation is completely different. \\

\textit{Discussion.} Equations \eqref{Section:Theory:Eq:Divergence} mean that the density of the matter-energy is \textit{dissipated}. This feature has been described through different settings, see \cite{Perez17,Perez18}, but the covariant expressions \eqref{Section:Theory:Eq:Divergence} in this paper are more general as they encode potential longitudinal modes which are forbidden in \cite{Perez17,Perez18} because of the diffeomorphism symmetry. Notice that we do not break diffeomorphism symmetry in extended GR, but instead, we introduce a novel source of energy (stress of the continuum describing dark energy) which also deforms the geodesic and trajectories of matter-energy, as described by equations \eqref{Eq:DefGeod:General} and by postulate (P3). Thus, the source of the violation of standard energy conservation and the possible existence of longitudinal modes are inherited from the reciprocal deformation of the medium and trajectories of the matter-energy. Notice that the combination of postulates (P2) and (P3) makes the extended GR theory consistent, complete, and constitutive, see Figure \ref{Figure:DiagRepresentationTheory}.

\begin{figure}
\begin{tikzpicture}[every text node part/.style={align=center},scale=0.7]
    \node (Continuum) [ellipse, draw,
    minimum width = 2cm, 
    minimum height = 1.5cm, fill=white] at (-6,-4) {Extended theory \\ of General relativity};  
    \node (Space) [ellipse, draw,
    minimum width = 1cm, 
    minimum height = 1.5cm, fill=white] at (-1.5,0.75) {Metric: $g_{\mu\nu}$};
    \node (Stress) [ellipse, draw,
    minimum width = 2cm, 
    minimum height = 1.5cm, fill=white] at (0.5,-4) {Stress: $\sigma_{\mu\nu}$};
    \node (Strain) [ellipse, draw,
    minimum width = 2cm, 
    minimum height = 1.5cm, fill=white] at (-1.1,-9) {Deformation: $G_{\mu}$ \\ strain: $\varepsilon_{\mu\nu}$};
    \node (Matter) [ellipse, draw,
    minimum width = 2cm, 
    minimum height = 1.5cm, fill=white] at (7.2,-4) {Matter-energy: $T_{\mu\nu}$};
    \draw [-latex] (Continuum) -- node [text width=2.5cm,midway,above,align=center,rotate=45 ] {\small \textbf{Geometry \\ space-time}} (Space);
    \draw [-latex] (Continuum) -- node [text width=2.5cm,midway,above,align=center,rotate=-45 ] {\small \textbf{Continuous \\ medium}} (Strain);
    \draw [latex-latex] (Space) -- node [text width=2.5cm,midway,above,align=center,rotate=-30 ] {\small Eqs. \eqref{Eq:Action:Modified}-\eqref{Section:Theory:Eq:ModEinstein}} (Matter);
    \draw [latex-latex] (Space) -- node [text width=2.5cm,midway,above,align=center,rotate=-65 ] {\small Eqs. \eqref{Eq:Action:Modified}-\eqref{Section:Theory:Eq:ModEinstein}}  (Stress);
    \draw [latex-latex] (Stress) -- node [text width=2.5cm,midway,above,align=center,rotate=0 ] {\small Eq. \eqref{Section:Theory:Eq:Divergence}} (Matter);
    \draw [latex-latex] (Stress) -- node [text width=2.5cm,midway,above,align=center,rotate=70 ] {\small Eq. \eqref{Section:Theory:Eq:StressTensor1}} (Strain);
    \draw [latex-latex] (Strain) -- node [text width=2.5cm,midway,above,align=center,rotate=33 ] {\small Eqs. \eqref{Eq:LangrangianInteractionDefMatter}-\eqref{Eq:DefGeod:General}} (Matter);
\end{tikzpicture}\caption{\textbf{Diagrammatic representation of the theory.} Our Extended theory of general relativity features both geometrical (metric $g_{\mu\nu}$) and physical (deformation $G_\mu$, strain $\varepsilon_{\mu\nu}$) aspects. These two aspects are interconnected through a consistent mathematical framework. }\label{Figure:DiagRepresentationTheory}
\end{figure}

\section{Cosmological expansion: revisited.}\label{Section:Cosmo} 

\subsection{Modified $\Lambda$CDM model}\label{Section:LambdaCDM}

\textit{Isotropic deformation of a 3+1D medium.} Now, we consider the Friedmann-Lema\^{i}tre-Robertson-Walker (FLRW) metric $$
ds^2 = -c^2dt^2 + a(t)^2\left(\frac{dr^2}{1-kr^2}+r^2 d\theta^2+r^2 \sin^2(\theta)d\phi^2\right)
$$ 
where $k$ is the scalar curvature. Recent observations support the hypothesis of zero curvature $\kappa=0$ \cite{WMAP,Planck16}, while other observations favor the closed universe assumption with $k>0$ \cite{Valentino20}. From the cosmological principle (i.e., assuming the medium is homogeneous and isotropic), we find that the stress tensor can be expressed as 
\begin{equation}\label{Section:LambdaCDM:sigma}
\sigma_{\mu\nu}=B\varepsilon(t) g_{\mu\nu} +3S\left(\varepsilon_{\mu\nu}-\frac{1}{4}\e(t) g_{\mu\nu}\right)
\end{equation}
where $B$ is the bulk modulus, $S$ is the shear modulus, and $\varepsilon(t)\equiv\varepsilon_\mu^\mu$ is the rate of deformation. 
If we assume that $\varepsilon_{\mu\nu}$ is invariant by rotation (no shear deformations/strains), we have $\varepsilon_{ij} = 0,\ \text{for}\ i\neq j$. In theory, we could have time-space shear deformations $\varepsilon_{0i}\neq 0,\ i\neq 0 $, but we assume they are equal to zero. This means that the 4-vector deformation $G_\mu$ is a function of time for $\mu=0$ and of the variable $r$ for $\mu\neq 0$. This is a reasonable assumption for the scaling factor $a(t)$ is stretched independently of the location in space and vice et versa. In this case, we obtain that $\epsilon = \text{Tr}\left(\varepsilon_{\mu\nu}\right) = \varepsilon_\mu^\mu = \alpha + 3\beta$. The components of the stress tensor are 
\begin{equation}\label{Section:LambdaCDM:sigmaComp0}
    \begin{cases}
    \sigma_0^0 = \left(B+\frac{9}{4}S\right)\alpha + \left(3B - \frac{3}{4}S\right)\beta = B\varepsilon +\frac{9S}{4}\Delta \ , \\
    \sigma_i^i = \left(B-\frac{3}{4}S\right)\alpha + \left(3B+\frac{3}{4}S\right)\beta = B\varepsilon - \frac{3S}{4}\Delta \ ,\ i\neq 0\ , \\
    \sigma_\mu^\nu = 0 \ ,\ \mu\neq \nu\ ,
    \end{cases}
\end{equation} 
where we recall that $\epsilon \equiv \epsilon_\mu^\mu$ is the trace of the strain tensor and $\Delta \equiv \alpha - \beta$ is the space-time anisotropy deformation rate.
Thus, we find that the trace of the stress and strain tensors are proportional 
\begin{equation}\label{Section:LambdaCDM:B}
\sigma = \sigma_{\mu}^\mu = 4B\epsilon ,  
\end{equation}
which is similar to a three-dimensional elastic medium $\sigma_i^i = 3B \epsilon_i^i$.

Finally, we find the equation of state
\begin{equation}\label{Section:LambdaCDM:EqState0} 
P = w \rho c^2\ ,
\end{equation}
where $P \equiv +\sigma_{i}^{i},\ i=,1,2,3$ and $\rho c^2 \equiv -\sigma_0^0$ (because of the signature $(-+++)$), with
\begin{equation}\label{Section:LambdaCDM:w0} 
w = -\frac{\left(B-\frac{3S}{4}\right)\alpha + \left(3B+\frac{3S}{4}\right)\beta}{\left(B+\frac{9S}{4}\right)\alpha + \left(3B-\frac{9S}{4}\right)\beta} = -\frac{4B\epsilon - 3S\Delta}{4B\epsilon + 9S\Delta}\ .    
\end{equation}
Notice that in general the coefficient $w$ is time-dependent and that $w=-1$ (dark energy-like) is either $S=0$ or $\Delta = 0$, see Table~1, and more specifically the isotropic deformation and shearless stress rows, respectively. If $\epsilon = 0$ or $B=0$, we find $w=1/3$ (radiation-like), which is expected as in this case $\sigma = 0$. If $\beta = 0$ and $S=4B/3$, we find $w=0$ (cold matter-like), i.e., the pressure is negligible. The speed of sound $c_s$ is determined as
\begin{equation}\label{Section:LambdaCDM:cs0} 
c_s = \sqrt{\frac{|P|}{\rho}} = c\sqrt{|w|}\ ,
\end{equation}
which is less or equal than $c$ if $|w|\leq 1$. Interestingly, for $w=-1$ (dark energy-like) the speed of sound equals the speed of light. This means that the longitudinal waves propagate at the same speed as the transversal waves. For $w=1/3$ the sound waves $c_s = c/\sqrt{3}$, which is analog to the cosmic sound \cite{CosmicSound05}.  \\

\begin{table}[h!]
\begin{tabular}{|p{2.4cm}||p{1.1cm}|p{1.1cm}|p{1.1cm}|p{1.4cm}|p{1.1cm}|p{1.3cm}||}
\hline
 & \vfill $\beta$ \vfill & \vfill $\Delta$ \vfill  & \vfill $\varepsilon$ \vfill & \vfill $B$ \vfill  & \vfill $S$ \vfill  & \vfill $w$ \vfill \\ 
 
\hline\textbf{\vfill Incompressible deformation \vfill } & \vfill ANY \vfill  & \vfill ANY \vfill & \vfill $0$ \vfill  & \vfill $\infty$  \vfill $B\varepsilon\rightarrow cst$ \vfill & \vfill ANY \vfill & \vfill $\geq -1$ \vfill \\

  \hline\textbf{\vfill Isotropic Deformation \vfill} & \vfill ANY \vfill  & \vfill $0$ \vfill & \vfill ANY \vfill  & \vfill ANY \vfill & \vfill ANY \vfill & \vfill $-1$ \vfill \\
  
 \hline\textbf{\vfill Uniaxial strain \vfill}  & \vfill $0$ \vfill  & \vfill ANY \vfill & \vfill ANY \vfill  & \vfill ANY \vfill & \vfill ANY \vfill & \vfill $-\frac{B-3S/4}{B+3S/4}$ \vfill \\
 
  \hline\textbf{\vfill Shearless stress \vfill}  & \vfill ANY \vfill  & \vfill ANY \vfill & \vfill ANY \vfill  & \vfill ANY$\neq 0$ \vfill & \vfill $0$  \vfill & \vfill $-1$ \vfill \\
  
\hline
\end{tabular}
\caption{\textbf{Elastic deformation regimes and corresponding equations of state.}
Summary of representative deformation and stress configurations in the elastic-medium
framework, characterized by the parameters $\beta$, $\Delta$, $\varepsilon$, bulk
modulus $B$, shear modulus $S$, and the resulting effective equation-of-state
parameter $w$. The table highlights several limiting cases, including incompressible
deformation, isotropic deformation, uniaxial strain, and shearless stress.
In particular, for an isobaric configuration with isotropic stress
($P_1=P_2=P_3=-\rho c^2$, corresponding to $w=-1$), the elastic medium admits
distinct regimes such as incompressible deformation, isotropic auxetic deformation,
and uniaxial deformation, depending on the relative values of $B$, $S$, and the
deformation parameters.}\label{Table1}
\end{table}

It is clear that the isotropic deformation or shearless stress, see Table \ref{Table1} match the interpretation of the $\Lambda$CDM model. However, the theory proposes a different mathematical and physical description of the dark energy and shows two distinct situations depending on the sign of the bulk modulus. 
From Eqs. \eqref{Section:Theory:Eq:ModEinstein}-\eqref{Section:LambdaCDM:sigma}, we derive the set of modified Friedmann equations describing the general dynamics of the cosmological medium and of the matter-energy in the FLRW metric space:
\begin{subequations}\label{Section:LambdaCDM:Eq:ModFriedmann0} 
\begin{equation}\label{Section:LambdaCDM:ModFriedmann00} 
\frac{\dot{a}(t)^2}{a(t)^2}+\frac{kc^2}{a^2}=\frac{8\pi G}{3 c^2}\left(\rho(t) c^2 - B \epsilon(t) - \frac{9S}{4}\Delta(t)\right)\ ,
\end{equation}
\begin{equation}\label{Section:LambdaCDM:ModFriedmannrr} 
\frac{\ddot{a}(t)}{a(t)}=-\frac{4\pi G}{3 c^2}\left(\rho(t) c^2 + 3 p + 2 B \epsilon(t) - \frac{9S}{2}\Delta(t)\right)\ .
\end{equation}
\end{subequations}
From these two equations, we find the following identity:
\begin{equation}\label{Section:LambdaCDM:ModFriedmannDiv0} 
\dot{\rho}+3\frac{\dot{a}}{a}\left(\rho+\frac{P}{c^2}\right)=\frac{B}{c^2}\dot{\varepsilon}+\frac{9S}{4c^2}\dot{\Delta}+9\frac{\dot{a}}{a} \frac{S}{c^2}\Delta\ .
\end{equation}
In our model, we consider that the strain tensor also deforms the geodesic, see equation \eqref{Eq:DefGeod:General}. It follows that for any matter-energy contained in the space, the modified equations of motion read:
\begin{equation}\label{Section:LambdaCDM:ModFriedmann3:v0} 
\dot{\rho}^{(j)}+3\frac{\dot{a}}{a}\left(\rho^{(j)}+\frac{P^{(j)}}{c^2}\right)=-\alpha\dot{\rho}^{(j)}-\frac{3}{a^2}\left(\frac{\dot{\beta}}{2}-\frac{\dot{a}}{a}\beta\right)\left(\rho^{(j)}+\frac{P^{(j)}}{c^2}\right)\ ,
\end{equation}
where $\rho^{(j)},\ j=1,2,3,\cdots$ are the densities of matter-energy (e.g., cold matter, electromagnetic radiation, etc..) and the $P^{(j)}$'s are the associated pressures, $\rho = \sum_j \rho^{(j)}$ and $P = \sum_j P^{(j)}$, and where $\alpha=\alpha(t)=\varepsilon_0^0(t)$ and $\beta=\beta(t)=\varepsilon_i^i(t),\ i\neq 0$.

\subsection{Stress effort in the time-direction and Poisson ratio} 

We can interpret the 3+1D elastic deformation as follows. Consider that the 3+1D volume $\mathcal{V} = x^0\prod_{i=1}^{3} x^i$ is stretched along the time-direction:
$$
x^0\mapsto x^0+\delta x^0 = x^0 + \varepsilon_0^0 x^0\ ,
$$
where $\varepsilon_0^0 = \frac{\delta x^0}{x^0}$ is the rate of deformation along the time-direction. Here, we consider small deformations, i.e., $|\delta x^0|\ll |x^0|$ $\Leftrightarrow$ $\varepsilon_0^0 \ll 1$. 
While stretched along the time-direction, the volume contract or expand isotropically in the other spatial directions $x^i \mapsto x^i +\delta x^i,\ i=1,2,3$:
$$
\varepsilon_i^i \equiv \frac{\delta x^i}{x^i} = -\nu \frac{\delta x^0}{x^0} -\nu\varepsilon_0^0 \ ,\ i=1,2, ,
$$
where $\nu$ is the Poisson ratio and $\varepsilon_i^i$ is the rate of deformation along the spatial directions. It turns out that the 3+1 volume  
\begin{align*}
\mathcal{V}' &= (x^0 +\delta x^0) \prod_{i=1}^3 \left(x^i + \delta^i \right) \\
&=  (x^0 +\varepsilon_0^0 x^0) \prod_{i=1}^3 \left(x^i +\varepsilon_i^i x^i\right) \\
&\approx \varepsilon_0^0 x^0\prod_{i=1}^3 x^i +x^0 \sum_{i=1}^{3} \varepsilon_i^i x^i\ \prod_{j\neq i, j\neq 0}x^j\ ,
\end{align*}
where we only kept the first order terms in $\varepsilon_0^0$. It follows that the rate of change of the 3+1D volume 
\begin{equation}\label{Section:LambdaCDM:epsilonnu}
    \frac{\delta \mathcal{V}}{\mathcal{V}} \approx \varepsilon =  \left(1-3\nu\right)  \epsilon_0^0
\end{equation}
as $\varepsilon = \varepsilon_0^0 + \sum_{i=1}^{3}\varepsilon_i^i = \left(1-3\nu\right)  \epsilon_0^0$. 
This equation is the 3+1D analog of the deformation rate for a three-dimensional continuous medium stretched in one direction (e.g., in the direction $x$): $\delta V/V \approx \varepsilon_x^x (1-2\nu)$. 
For $\beta =\varepsilon_i^i,\ i=1,2,3$, and $\varepsilon_i^i = -\nu \varepsilon_0^0$, we find that $\Delta =\alpha-\beta = (1+\nu)\varepsilon_0^0 = (1+\nu)\alpha$. Hence, we can rewrite equation \eqref{Section:LambdaCDM:sigmaComp0} as
\begin{equation}\label{Section:LambdaCDM:sigmaComp02}
    \begin{cases}
    \sigma_0^0 =  B\alpha (1-3\nu) +\frac{9S}{4}\alpha (1+\nu) \ , \\
    \sigma_i^i = B\alpha (1-3\nu) -\frac{3S}{4}\alpha (1+\nu) \ ,\ i\neq 0\ , \\
    \sigma_\mu^\nu = 0 \ ,\ \mu\neq \nu\ ,
    \end{cases}
\end{equation}

The Young modulus is defined as the ratio between the strain and the stress in the direction of the deformation, whence
\begin{equation}\label{Section:LambdaCDM:YoungModulusE}
E = \frac{\sigma_0^0}{\varepsilon_0^0}\ .
\end{equation}
Then, considering an isotropic medium and uniaxial stress in the time direction, we shall have
\begin{equation}\label{Section:LambdaCDM:YoungModulusEdirect}
    \begin{cases}
    \sigma_0^0 = E \varepsilon_0^0 \ , \\
    \sigma_i^i = 0 \ ,\ i\neq 0\ .
    \end{cases}
\end{equation}
Therefore, the traces of the strain tensor ($\sigma$) is proportional to the stress $\alpha$:
\begin{equation}\label{Section:LambdaCDM:Eiso}
\sigma = 4B \epsilon = E\varepsilon_0^0 = E\alpha\ .
\end{equation}
After combining equations \eqref{Section:LambdaCDM:B}, \eqref{Section:LambdaCDM:epsilonnu} with \eqref{Section:LambdaCDM:Eiso}, we find the relation between the Bulk modulus $B$ and the Young modulus $E$
\begin{equation}\label{Section:LambdaCDM:KvsE}
    B=\frac{1}{4}\frac{E}{1-3\nu}\ .
\end{equation}
From equations \eqref{Section:LambdaCDM:sigmaComp02} and \eqref{Section:LambdaCDM:KvsE} 
we can easily deduce the relation between the shear modulus $S$ and the Young modulus $E$
\begin{equation}\label{Section:LambdaCDM:SvsE}
    S=\frac{1}{3}\frac{E}{1+\nu}\ .
\end{equation}
Notice the analogy with the three-dimensional relations $B=E/(3(1-2\nu))$ and $S = E/(2(1+\nu))$. 

The Poisson ratio determines the nature of the elastic deformation. From the relations \eqref{Section:LambdaCDM:KvsE}-\eqref{Section:LambdaCDM:SvsE}, it is clear that $-1 < \nu< 1/3$. 
There are three elastic regimes:\\
\begin{itemize}
    \item \textbf{Standard (S):\ } $0<\nu\leq 1/3$. In this case, the deformation rates in the spatial directions are negative (positive, resp.) $\varepsilon_i^i = -\nu \epsilon_0^0$ if the deformation rate in the time direction is positive (negative, resp.). This situation is standard in material science for three-dimensional materials. 
    \item \textbf{Auxetic (A):\ } $-1\leq\nu<0$. This case is the exact opposite of the previous one, i.e., the deformation rates in the spatial directions are positive (negative, resp.) $\epsilon_i^i = -\nu \varepsilon_0^0$ if the deformation rate in the time-direction is positive (negative, resp.). This situation has been investigated in material science for three-dimensional materials since the 90's \cite{Evans91}.  
    \item \textbf{Uniaxial deformation (UD):\ } $\nu=0$. In this situation, there is no deformation in the spatial directions $\varepsilon_i^i=0$. We say that space is \textit{rigid} in the sense that stretch or contraction in the time direction does not affect space directions. The three-dimensional analog is cork material for which the Poisson ratio is about $0$ (this makes cork well-suited as a material for wine bottle stoppers).  \\
\end{itemize}
In particular, we have two critical regimes when $\nu = 1/3$ (Incompressible regime), in which the 3+1D volume is unchanged (even though the spatial volume is compressed/stretched for $\alpha>0/\alpha<0$, respectively), and $\nu=-1$ (Isotropic Auxetic regime), in which the rate of deformation is the same for all directions (time and space). We summarized these particular regimes in Table \ref{Table2}. 
In what's next, we assume that there is no deformation of the space directions, i.e., the Poisson ratio $\nu=0$ that corresponds to the Uniaxial Deformation regime (UD), see Table \ref{Table2}. 
Hence, the components of the stress tensor now read 
\begin{equation}\label{Section:LambdaCDM:sigmaComp}
    \begin{cases}
    \rho c^2 = \sigma_0^0 = \left(B+\frac{9}{4}S\right)\epsilon\ , \\
    P = -\sigma_i^i = -\left(B-\frac{3}{4}S\right)\epsilon\ ,\ i\neq 0\ .
    \end{cases}
\end{equation} 
\begin{equation}\label{Section:LambdaCDM:w} 
w = -\frac{4B-3S}{4B+9S}    
\end{equation}

\begin{table}[h!]
\begin{tabular}{|p{2.9cm}||p{1.8cm}|p{1.8cm}|p{1.8cm}|p{1.8cm}||}
\hline
   & \vfill $\nu$ \vfill & \vfill $\epsilon$ \vfill  & \vfill $B/E$ \vfill & \vfill $S/E$ \vfill \\ 
\hline\textbf{\vfill Incompressible deformation (ID) \vfill } & \vfill $1/3$ \vfill  & \vfill $0$ \vfill & \vfill $\infty$ \vfill  & \vfill $1/4$ \vfill \\
 
  \hline\textbf{\vfill Isotropic Auxetic Deformation (IAD) \vfill} & \vfill $-1$ \vfill  & \vfill $4\alpha$ \vfill & \vfill $1/16$ \vfill  & \vfill $\infty$ \vfill \\

 \hline\textbf{\vfill Uniaxial Deformation (UAD) \vfill}  & \vfill $0$ \vfill  & \vfill $\alpha$ \vfill & \vfill $1/4$ \vfill  & \vfill $1/3$ \vfill \\
\hline
\end{tabular}
\caption{\textbf{Stress in the time-direction.} We consider stress only in the time direction (hence $P_i=0,\ i=1,2,3$) that describes cold matter state $w=0$. We find three different critical regimes: incompressible deformation (ID), isotropic auxetic deformation (IAD), and uniaxial deformation (UAD). }\label{Table2}
\end{table}

\subsection{Exact solutions to the modified Friedmann equations for the isobaric pressure model}\label{Section:ModFriedmannEq}

From equation \eqref{Section:LambdaCDM:w}, we find that the equation of states describes a negative pressure similar to that of vacuum energy for $w=-1$ if $S=0$ (no shear modulus) or if $\Delta =0 $ (isotropic deformation). Either way,  we find that the expression of the stress tensor of the cosmological medium is 
$$
\sigma_{\mu\nu}=B\varepsilon(t)g_{\mu\nu}\ .
$$
We straightforwardly notice the analogy with any time-dependent cosmological constant models with $\Lambda(t) \equiv \frac{8\pi G}{c^2}K\varepsilon(t)$. We introduced the constant $K=-B$ that can be interpreted as a stress-energy density that characterizes the medium for $K>0$. We find different scenarios depending on the signs of the bulk modulus $B=-K$ and the deformation rate $\epsilon$.\\
\begin{enumerate}
    \item[] \textbf{Case (i):\ }For $K>0$ ($B<0$) and $\epsilon>0$ the term $\sigma_{00}>0$ (i.e., $\Lambda(t)>0$) can be interpreted as the dark energy density corresponding to the stress-energy of the medium. The acceleration of the expansion of the cosmological medium is due to the presence of internal pressure (the spatial terms $\sigma_{ii}$ are negative) with a positive deformation rate ($\epsilon>0$).\\
    \item[] \textbf{Case (ii):\ } Similarly, in this case, the bulk modulus is positive ($K<0$ or $B>0$) while the deformation rate is negative $\epsilon<0$, we find that the energy density is also positive $\sigma_{00}>0$ and that the pressure is negative.\\
\end{enumerate}
The last two cases describe an elastic material experiencing positive pressure and correspond to a universe with decelerated expansion ($\Lambda<0$) : \\
\begin{enumerate}
    \item[] \textbf{Case (iii):\ } For $K>0$ with $\epsilon <0$ (negative bulk modulus and deformation rate).\\
    \item[] \textbf{Case (iv):\ } For $K<0$ with $\epsilon>0$ (positive bulk modulus and deformation rate). \\
\end{enumerate}

In what follows, we will assume that $\beta = 0 $ (homogeneity of space) and $S = 0$ (equation of state $w=-1$). From equations \eqref{Section:LambdaCDM:Eq:ModFriedmann0}, we find that the Friedmann equations read
\begin{subequations}\label{Section:LambdaCDM:Eq:ModFriedmann} 
\begin{equation}\label{Section:LambdaCDM:ModFriedmann1} 
H(t)^2 \equiv \frac{\dot{a}(t)^2}{a(t)^2}=\frac{8\pi G}{3 c^2}\left(\rho(t) c^2 + K\epsilon(t)\right)\ ,
\end{equation}
\begin{equation}\label{Section:LambdaCDM:ModFriedmann2} 
\sum_{w} \left(\dot{\rho}_{w}(t)+3(1+w)H(t)\rho_{w}(t) \right)=-\frac{K}{c^2}\dot{\varepsilon}(t)\ ,
\end{equation}
\begin{equation}\label{Section:LambdaCDM:ModFriedmann3} 
\dot{\rho}_{w}(t)+3(1+w)H(t)\rho_{w}(t)=-\varepsilon(t)\dot{\rho}_{w}(t)\ ,
\end{equation}
\end{subequations}
where $H(t) \equiv \dot{a}(t)/a(t)$ is the Hubble parameter, $\rho_w c^2$ stands for energy density of the pressureless cold matter for $w=0$, and of the radiation for $w=1/3$.   
Equation \eqref{Section:LambdaCDM:ModFriedmann2} has been derived from equation \eqref{Section:Theory:Eq:Divergence}, after taking $S=0$ in equation \eqref{Section:LambdaCDM:sigma} (which means that $w=-1$). Equation \eqref{Section:LambdaCDM:ModFriedmann3} is a consequence of the deformations of the equations of motion, see equation \eqref{Section:LambdaCDM:ModFriedmann3:v0} \footnote{See also the analogy with the geodesic deformation \eqref{Eq:GeoDefg} with $g_{00}\mapsto g_{00}+\e_{00}=1+\e(t)$.}. Equations \eqref{Section:LambdaCDM:ModFriedmann2} and \eqref{Section:LambdaCDM:ModFriedmann3} lead to 
$$
\varepsilon(t)=\varepsilon_0 \exp{\left\{\frac{c^2}{K}\left(\rho(t)-\rho_0\right)\right\}}
$$
where $\varepsilon_0\equiv \varepsilon(t_0)$ is the deformation rate at the present epoch $t_0$. This means that the deformation rate increases (decreases) with the density of matter $\rho(t)$ in case (i) (case (ii), respectively). As mentioned above, one can interpret the density  $\rho_{\Lambda}(t)c^2$ as the stress-energy density of the cosmological medium which is equal to the opposite bulk modulus $K=-B$ times the deformation rate $\e(t)$, see  equation \eqref{Section:LambdaCDM:ModFriedmann1}
\begin{equation}\label{Section:LambdaCDM:stress-energy-density} 
\rho_{\Lambda}(t)c^2 = K \varepsilon(t) = K \varepsilon_0 \cdot\exp{\left\{\frac{c^2}{K}\left(\rho(t)-\rho_0\right)\right\}}
\end{equation}
It follows that the cosmological constant $\Lambda = 10^{-52} m^{-2}$ corresponds to the stress-energy density at the present epoch multiplied by the constant $8\pi G/c^4$, i.e.,  $\Lambda = 8\pi G K\varepsilon_0/c^4$. 
Assuming that the universe will expand indefinitely, we find that $\rho_\Lambda(\infty) =  K\varepsilon_0 \exp{\left\{-\frac{c^2}{K}\rho_0\right\}}$. This scenario corresponds to a de Sitter model of infinite expansion with a smaller or bigger rate than the one predicted by the $\Lambda$CDM model for case (i) and case (ii), respectively.
In contrast with a constant cosmological constant in the $\Lambda$CDM model, the time variation of the stress energy density of the cosmological medium also changes the relation between the scaling factor and the total density of matter-energy in the matter-dominated ($w=0$) and radiation-dominated eras ($w=1/3$). From Eqs. \eqref{Section:LambdaCDM:ModFriedmann2} and \eqref{Section:LambdaCDM:ModFriedmann3}, we obtain
\begin{equation}\label{Section:LambdaCDM:scalingfactor} 
a(t)=\left(\frac{\rho_0}{\rho}\right)^{\frac{1}{3(1+w)}} \cdot \exp\left\{ \frac{\e_\infty}{3(1+w)}\left(\text{Ei}\left(\frac{c^2}{K}\rho_0\right)-\text{Ei}\left(\frac{c^2}{K}\rho\right)\right) \right\}\ ,
\end{equation}
where $\e_\infty \equiv \varepsilon_0 \exp{\left\{-\frac{c^2}{K}\rho_0\right\}}$ is the value of the deformation rate in the limit $t\rightarrow +\infty$. 
In Eq. \eqref{Section:LambdaCDM:scalingfactor}, the first algebraic term corresponds to the scaling factor for the standard $\Lambda$CDM and the term in the exponent is the modification caused by the deformation of the medium, where the exponential-integral function $\text{Ei}(x)\equiv \int_{-\infty}^{x} du\ u^{-1}\exp(u)$ \cite{Abr}. From Eq. \eqref{Section:LambdaCDM:scalingfactor} it is clear that the scaling factor can grow fast as the density decreases for case (i) while it asymptotically behaves as $\rho^{-3(1+w)}$ for case (ii). 

We shall now discuss the relation to the $\Lambda$CDM model. From Eqs. \eqref{Section:LambdaCDM:Eq:ModFriedmann}-\eqref{Section:LambdaCDM:scalingfactor} , it is now clear that we find the $\Lambda$CDM model in the limit $|K|\rightarrow +\infty$ and $|\e_0|\rightarrow 0$ and keeping the product constant $K\e_0 \rightarrow 3\Lambda c^2/(8\pi G)$, where the cosmological constant $\Lambda \sim 10^{-52}m^{-2}$. However, this limit is not valid if the total density of matter-energy is comparable with $K/c^2$ as the term in the exponential in Eq. \eqref{Section:LambdaCDM:stress-energy-density}  and Eq. \eqref{Section:LambdaCDM:scalingfactor} is no longer negligible.
Note that the extension of the result in the non-linear regime is straightforward for the FLRW model. Indeed, consider non-linear deformations for the variable $x^0=ct$, i.e., $ct\mapsto ct + G(t)$ where $G(t)$ is the non-zero- value of the deformation vector $G^\mu(x)$ corresponding to the time component $\mu=0$. Following a similar approach to the linear case, we find that $cdt\mapsto cdt + \dot{G}dt $ and so that  $c^2dt^2\mapsto \left(1 +2 \frac{\dot{G}}{c} + \frac{\dot{G}^2}{c^2}\right)c^2dt^2 $. This leads to the deformed metric $g_{00}\mapsto 1+ \e(t)$, with $\e(t) = 2 \frac{\dot{G}}{c} + \frac{\dot{G}^2}{c^2}$ instead of $2 \frac{\dot{G}}{c}$ for the linear case. Assuming Hooke's law, see Eq. \eqref{Section:Theory:Eq:StressTensor1}, we find the exact same equations as Eqs. \eqref{Section:LambdaCDM:Eq:ModFriedmann} - \eqref{Section:LambdaCDM:scalingfactor}. This yields some interesting results for case (i) which we are going to discuss next.


\begin{figure}
 \includegraphics[width= 0.47\columnwidth]{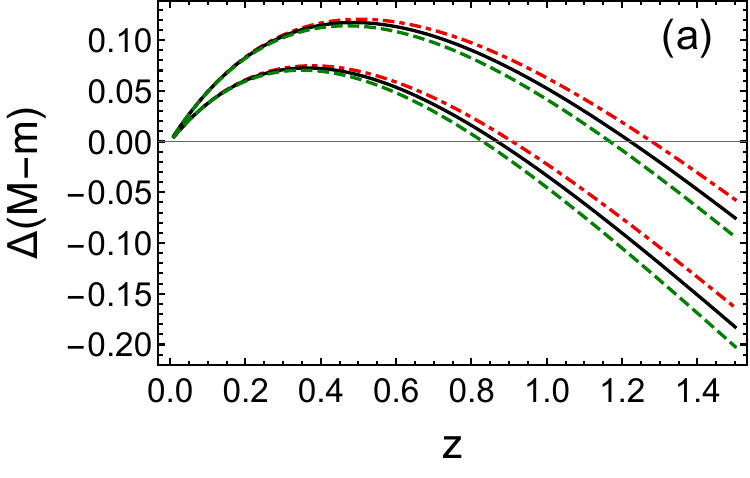} \includegraphics[width= 0.47\columnwidth]{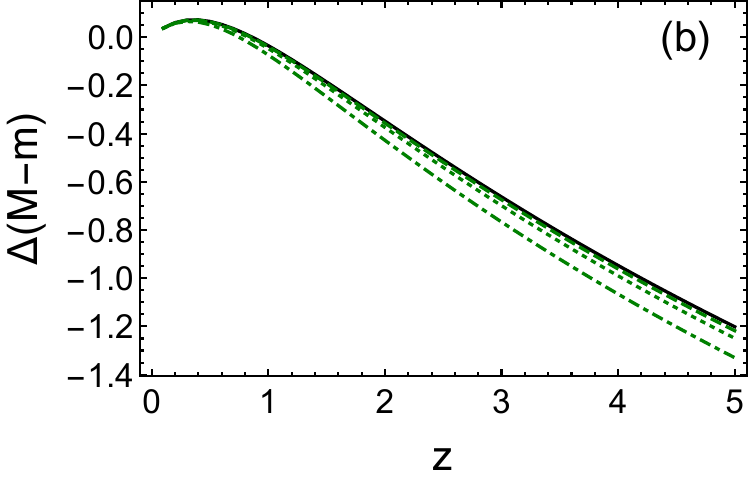}
\caption{{\bf Parameter estimation and relative distance modulus curves. }
In (a) we display relative distance modulus $\Delta(M-m)$ as a function of the redshift $z$ of the $\Lambda$CDM model (continuous curves) and of the modified $\Lambda$CDM model with the deformation rate at the present epoch $|\e_0|=10^{-1.5}$ (dotted-dashed lines for case (i) and dashed lines for case (ii)), for a dark energy density ratio $\Omega_{\Lambda}=0.692$, \cite{Planck16} (three curves at the top) and $\Omega=0.62$, \cite{Hubble16} (three curves at the bottom). This shows that for a given dark energy density, the curves for case (ii) are below the $\Lambda$CDM model, while the curves for case (i) are above. In (b) we show $\Delta(M-m)$ only for $\Omega_{\Lambda}=0.62$, \cite{Hubble16} for the $\Lambda$CDM model (continuous curve) and for case (ii) with $|\e_0| = 10^{-2};\ 10^{-1.5},\ 10^{-1}$ (dashed, dotted, dotted-dashed lines, respectively).  }\label{Fig1}
\end{figure}

\subsection{Cosmological expansion - Parameters estimation.}\label{Section:Expansion} 

In our model, we introduced two new fundamental parameters, namely the bulk modulus $B=-K$ and the deformation rate $\e_0$ at the present epoch. We propose to examine their effects on the luminosity curves of celestial bodies. We assume that the product between these two quantities equals the density of dark energy at the present time, e.g., see the values in \cite{Planck16} or in \cite{Hubble16}.  We recall that in case (i) the bulk modulus is negative (i.e., $K>0$) and the deformation rate is positive ($\e_0>0$) while in case (ii) the signs are the opposite. 
The luminosity distance of a celestial body distant with a redshift $z$ from an observer is given by \cite{Wright07}
$$
d_L(z) = c(1+z) \int_{0}^{z} \frac{dz'}{H(z')}\ ,
$$
where $H(z')$ is the Hubble factor depending on the redshift $z'$.   
In astronomy, it is useful to introduce the distance modulus \cite{Wright07}
$$
\text{m}-\text{M} = 5\log_{10}\left(\frac{d_L}{10 \text{pc}}\right)\ ,
$$
where the conversion of the parsec unit is $1\text{pc} = 3.086 \times 10^{16} \text{m}$.  
In Fig. \ref{Fig1}, we display relative distance modulus $\Delta(M-m)$ calculated by subtracting the distance modulus computed for the empty universe Milne model \cite{Wright07} from the distance modulus of the $\Lambda$CDM model and of the modified $\Lambda$CDM model.  Fig. \ref{Fig1}(a) we show $\Delta(M-m)$ for two different values of the density of dark energy ratio $\Omega_\Lambda\equiv \rho_\Lambda/\rho_c$, namely, $\Omega_\Lambda=0.692$ obtained by the Planck Collaboration \cite{Planck16} and $\Omega_\Lambda=0.62$ obtained by the Hubble Space Telescope \cite{Hubble16}, where the critical density $\rho_c \equiv3H(0)^2/(8\pi G)$. We can see that the relative modulus for case (i) (for case (ii)) is greater (smaller, respectively) than that of the $\Lambda$CDM for fixed $\Omega_\Lambda$. In Fig. \ref{Fig1}(b), we illustrate the variation of the relative distance modulus for case (ii) for different values of $\e_0$.  Fig. \ref{Fig1} shows qualitatively that the additional parameter $\e_0$ offers the possibility to find a better agreement between the Planck Collaboration and the Hubble Space Telescope \cite{Planck16,Hubble16}. We mention that it is possible to modify the effective equation of state for the dark energy by taking a non-zero shear modulus $S$, see Eq. \eqref{Section:LambdaCDM:sigma}. It is also possible to include an effective space-curvature $\kappa$ by adding a deformation rate in the radial coordinate. This is beyond the scope of this paper, but we hope this could motivate future research.


\section{Cosmological inflation: revisited.}\label{Section:Inflation}

\subsection{Inflation mechanism of an elastic medium}

Interestingly, case (i) (negative bulk modulus $K=-B>0$ and positive deformation
rate $\epsilon_0>0$) exhibits features closely analogous to standard inflationary
scenarios \cite{Starobinskii80,Guth81,Albrecht82,Linde82}. The quantity $c^2/K$
appearing in the exponential terms of
Eqs.~\eqref{Section:LambdaCDM:stress-energy-density} and
\eqref{Section:LambdaCDM:scalingfactor} has the dimension of the inverse of a
density, allowing one to define the characteristic density
\begin{equation}
\rho_d \equiv \frac{K}{c^2},
\end{equation}
which marks the threshold between two distinct dynamical regimes. In the first
regime, for $\rho\ll\rho_d$, elastic deformations are negligible and the scale
factor follows the standard behavior $a\sim\rho^{-1/(3+3w)}$. In the second
regime, for sufficiently large densities $\rho\gg\rho_d$, elastic effects become
dominant and the scale factor decreases super-exponentially as the density grows,
\begin{equation}\label{Eq:Infl:a(t)}
a(t)\sim a(t_1)\exp\!\left[-\frac{\rho_\Lambda(t)-\rho_{\Lambda}(t_1)}{4\rho_s}\right],
\end{equation}
as illustrated in Fig.~\ref{Fig2}.

In this high-density regime, the dark-energy density increases exponentially fast
once the matter-energy density reaches values of order
$\rho_s\sim 10^2\rho_d$, ensuring $\rho\gg\rho_d$. Consequently, for case (i), the
present-day deformation rate $\epsilon_0$ must be extremely small, corresponding
to a very large value of $\rho_d$ compared to the matter-energy density at the CMB
epoch, $\rho_{\mathrm{CMB}}\sim 10^{-18}\,\mathrm{kg\,m^{-3}}$.

{Although the elastic-medium framework formally involves two parameters, the bulk
modulus $K$ and the present-day deformation rate $\epsilon_0$, these are not
independent once observational constraints are imposed. The present-day dark
energy density fixes their product through
$K\epsilon_0=\rho_{\Lambda,0}c^2$, so that specifying $K$ uniquely determines
$\epsilon_0$. The value of $K$ is chosen so as to place the characteristic density
$\rho_d=K/c^2$ well above all post-inflationary energy densities, while remaining
below the Planck density. This ensures that elastic effects are negligible during
standard cosmological evolution, but become dominant in the high-density regime
relevant to the early Universe.}

We now consider an illustrative inflationary scenario. We fix the bulk modulus to
$K=1.64\times10^{109}\,\mathrm{N\,m^{-2}}$ and impose the observational
normalization $K\epsilon_0=\rho_\Lambda(0)c^2$, where
$\rho_\Lambda(0)=6.023\times10^{-27}\,\mathrm{kg\,m^{-3}}$ is the present-day dark
energy density (Hubble Space Telescope 2016 value \cite{Hubble16}). For this
choice, the characteristic density is
$\rho_d=1.82\times10^{92}\,\mathrm{kg\,m^{-3}}\simeq3.5\times10^{-4}\rho_{\mathrm{Pl}}$,
and the critical density $\rho_sc^2$ governing the onset of the super-exponential
regime is of order $10^{-3}\rho_{\mathrm{Pl}}$. Here
$\rho_{\mathrm{Pl}}c^2=E_{\mathrm{Pl}}^4/(\hbar c)^3\approx2\times10^{113}\,
\mathrm{J\,m^{-3}}$ is the quantum fluctuation (Planck) density, with
$E_{\mathrm{Pl}}\approx1.6\times10^{19}\,\mathrm{GeV}$.

When $\rho>\rho_d$, the dark-energy density grows exponentially and reaches the
Planck density $\rho_{\mathrm{P}}c^2\approx4.66\times10^{113}\,\mathrm{J\,m^{-3}}$,
as shown in Fig.~\ref{Fig2}(a). Correspondingly, the scale factor decreases from
$\sim10^{-32}$ to $\sim10^{-60}$ once $\rho>\rho_s\sim10^{-2}\rho_{\mathrm{Pl}}$,
see Fig.~\ref{Fig2}(b).

{To facilitate comparison with standard inflationary scenarios, we associate an effective inflationary energy scale with the
elastic stress--energy density driving the accelerated expansion,
\begin{equation}
E_{\mathrm{inf}}(t)\equiv\left[(K\epsilon(t))(\hbar c)^3\right]^{1/4}.
\end{equation}
For the illustrative parameter values considered here, this corresponds to a
high-scale inflationary regime with
$E_{\mathrm{inf}}\sim10^{18}$--$10^{19}\,\mathrm{GeV}$. Importantly, this scale is
not imposed a priori but emerges dynamically from the elastic evolution once the
late-time normalization is fixed.}

The duration of the inflationary phase can be estimated from
\begin{equation}
\Delta t=\int_{a_s}^{a_{\mathrm{Pl}}}\frac{da}{aH}
=\sqrt{\frac{3}{8\pi G\rho_s}}\int_{\rho_s}^{\rho_f}d\rho
\left(\frac{da}{d\rho}\right)\left[\rho+\rho_\Lambda(\rho)\right]^{-1/2}
\approx2.2\times10^{-42}\,\mathrm{s},
\end{equation}
where $\rho_s=9.5\times10^{-3}\rho_{\mathrm{Pl}}$ and
$\rho_f=10^{-2}\rho_{\mathrm{Pl}}$. A similar estimate yields
$t_{\mathrm{GU}}\sim10^{-38}\,\mathrm{s}$ for the epoch at which the energy density
reaches $\rho_{\mathrm{GU}}c^2=E_{\mathrm{GU}}^4/(\hbar c)^3$, corresponding to a
grand unification scale $E_{\mathrm{GU}}\sim10^{16}\,\mathrm{GeV}$.

\subsection{{Naturalness and initial conditions.}}

{An important question for any inflationary framework concerns the role of initial
conditions and the degree of fine-tuning required to trigger accelerated
expansion. In the elastic-medium model considered here, accelerated expansion does
not arise for arbitrary choices of parameters: in particular, it requires
appropriate signs of the bulk modulus $K$ and of the deformation rate
$\epsilon(t)$. Outside this sector, the dynamics lead instead to decelerated
expansion or contraction.}

{Within the physically relevant regime identified in this work, however, the
inflationary solutions display an attractor-like behavior. Owing to the
super-exponential dependence of the elastic stress--energy density on the matter
density, a broad range of initial deformation rates evolves toward an
inflationary phase without the need for precise adjustment of initial conditions.
An analogous mechanism governs the late-time dark-energy regime, where the
deformation rate relaxes dynamically to a small but nonvanishing value, yielding
accelerated expansion at cosmological scales.}

{A quantitative characterization of the full phase space of initial conditions,
including measures of basin size and stability, would require a dedicated
dynamical systems analysis and is left for future work. The present discussion is
therefore intended as a qualitative assessment, indicating that inflationary and
dark-energy phases do not rely on isolated or finely tuned initial states within
the elastic-medium framework.}

\subsection{Transition from the inflation-like to the dark-energy-like period}

Notice that the deformation rate $\varepsilon(t)$ also decreases super-exponentially fast, as it is proportional to the density $\rho_\Lambda(t)$. This implies that the continuous medium, initially modeling the inflaton field, transitions from being elastically stretched during the inflationary period to becoming effectively rigid afterward, thereby modeling dark energy. This is caused bythe large value of the bulk modulus $K \sim 10^{109} \text{N}\cdot\text{m}^{-2}$ and by to the super-fast expansion of the scaling factor, which causes the dilution of the deformation strength of the medium. 

By analogy with material science, we can say that during the inflation period, the cosmological medium experiences a phase transition between an elastic state (large value of the deformation rate $\varepsilon$ despite the large value of the bulk modulus $K$) and a solid-state (relatively small/large value of deformation rate/bulk modulus, respectively). We interpret that medium looks stiff and non-deformable in our present epoch by this very small value of the deformation rate $\varepsilon_0 \sim 10^{-133}$. In this limit, the dissipative terms in equation \eqref{Section:LambdaCDM:ModFriedmann2} become negligible ($\dot{\varepsilon} \sim 0$), and hence, the energy is conserved, as stated in the standard GR. Also, as $\varepsilon(t) = 2D_\mu G^\mu \sim 0$ in this limit, our model becomes analogous to that of \cite{Perez17}, which states the diffeomorphism symmetry.

 \begin{figure}[h!]
 \includegraphics[width= 0.49\columnwidth]{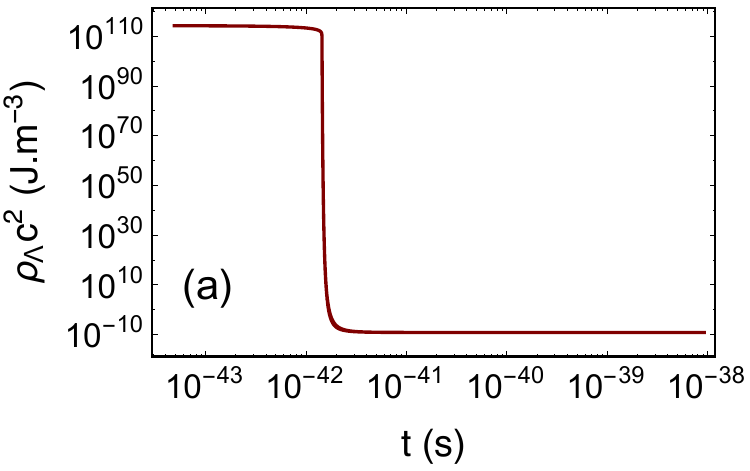} \includegraphics[width= 0.49\columnwidth]{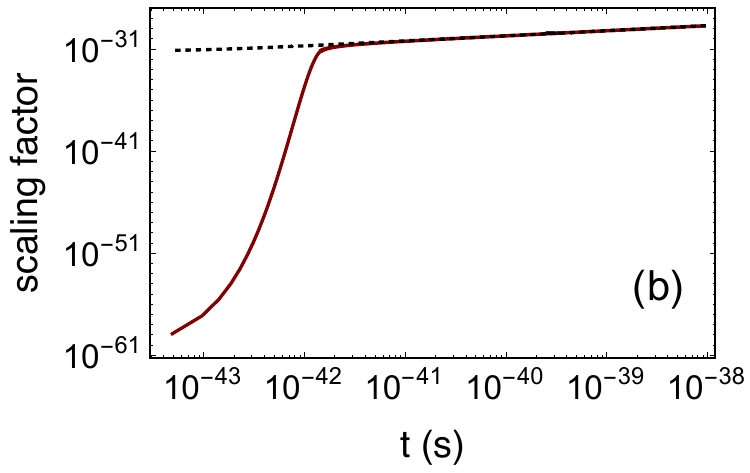} 
\caption{{\bf Inflation mechanism. }  In this figure we plot the dark energy density curve (a), the scaling factor (b) for the modified $\Lambda$CDM (continuous line) and the $\Lambda$CDM standard model (dashed line), as functions of time, for $K = 1.64\times 10^{109}\text{N}\cdot\text{m}^{-2}$, where $ \rho_{\Lambda}(0)=6.023\times 10^{-27}\text{kg}\cdot\text{m}^{-3}$ is the dark energy density at the present epoch (see, \cite{Hubble16}). These two graphs show the fast decay of the dark energy density and the super-exponential growth of the scaling factor. }\label{Fig2}
\end{figure}

\subsection{{Effective reheating via elastic--matter coupling}}

{A fundamental aspect of any inflationary scenario is the mechanism by which the
Universe transitions from an accelerated expansion phase to a hot,
radiation-dominated epoch. In standard scalar-field inflation, this transition is
achieved through reheating, i.e., the decay of the inflaton field into matter and
radiation \cite{LindeBook,MukhanovBook}. In the present framework, inflation is not
driven by a scalar field but by the elastic deformation of a four-dimensional
continuous medium. Reheating must therefore be reinterpreted accordingly.}

{In our model, the stress-energy tensor associated with elastic deformation,
$\sigma_{\mu\nu}$, is dynamically coupled to the matter content of the Universe.
As shown in equation \eqref{Section:Theory:Eq:Divergence}, the standard conservation law for matter-energy is modified
according to
\begin{equation*}
D_\nu T^{\mu\nu} = - D_\nu \sigma^{\mu\nu},
\end{equation*}
which expresses the exchange of energy and momentum between matter and the elastic
degrees of freedom of the cosmological medium. This relation provides a natural
and universal channel for energy transfer, independent of the microscopic nature
of the matter fields. Modified conservation laws leading to energy exchange
between vacuum-like components and matter have been widely explored in
cosmological contexts; see, e.g.,~\cite{Barrow1986,ShapiroSola2010,Perez17}.}

{During the inflationary phase, a significant fraction of the total energy density
is stored in the elastic deformation of the medium, characterized by a large
deformation rate $\epsilon(t)$. As the Universe undergoes super-exponential
expansion, the deformation rate $\epsilon(t)$ rapidly decreases (see Fig.~4),
leading to a decay of the elastic stress-energy density
\begin{equation*}
\rho_\Lambda(t)c^2 = K\,\epsilon(t).
\end{equation*}
Through the coupling encoded in equation \eqref{Section:Theory:Eq:Divergence}  and equation \eqref{Section:LambdaCDM:ModFriedmann2}, this decay acts as a source term for matter and radiation, effectively converting elastic energy into particle excitations.}

{From this perspective, reheating corresponds to the relaxation of the elastic
medium toward an effectively rigid state, accompanied by the production of
matter and radiation. This mechanism is conceptually analogous to reheating or
preheating in scalar-field models, but without invoking a specific inflaton
potential or decay channel. Instead, reheating emerges as a macroscopic
consequence of elastic energy dissipation into matter through a covariant and
universal coupling.}

{Although a precise determination of the reheating temperature requires specifying
the microscopic coupling between elastic degrees of freedom and matter fields, a
parametric estimate can be obtained. If a fraction $\eta$ of the elastic energy
density at the end of inflation, $u_{\rm end}=K\epsilon(t_{\rm end})$, is
converted into radiation, the reheating temperature follows from
$u_{\rm rad}=(\pi^2/30)g_*T_{\rm rh}^4$, yielding
\begin{equation}
T_{\rm rh}\simeq
\left(\frac{30}{\pi^2 g_*}\,\eta\,u_{\rm end}\right)^{1/4},
\end{equation}
where $g_*$ denotes the effective number of relativistic degrees of freedom at the
reheating temperature.
For $u_{\rm end}\sim10^{-3}u_{\rm Pl}$, $g_*\sim\mathcal{O}(10^2)$, and
$\eta\sim\mathcal{O}(1)$, this corresponds to a high reheating temperature
$T_{\rm rh}\sim10^{15}$--$10^{16}\,\mathrm{GeV}$.}

{A detailed microscopic description of particle production, including reheating
efficiency, thermalization timescales, and the resulting reheating temperature,
would require specifying the coupling between the deformation field and
particular matter fields. Such an analysis lies beyond the scope of the present
work. Here, we emphasize that the framework naturally accommodates reheating at
the phenomenological level and provides a consistent mechanism for repopulating
the Universe with matter and radiation following the inflationary phase. A
quantitative analysis of reheating efficiency and particle production within
this framework will be presented in future work.}

\section{Discussion and outlook}

Our model introduces two additional fundamental parameters, namely, the bulk modulus of the medium $B=-K$ and the deformation rate $\epsilon_0$ at the present epoch.  
We investigated different scenarios of expansion depending on the sign of the deformation rate and of the bulk modulus, either (i) $B<0$ and $\e_0>0$ or (ii) $B>0$ and $\e_0<0$, see Fig. \ref{Fig1}. This raises the question of the microscopic origin of the signs. We think that the values of the bulk modulus and the deformation rate could be derived from a more fundamental microscopic theory as it was done in material science, see \cite{Goldhirsch02}. Heuristically, we can interpret the negative sign of the deformation rate of the medium from cases (i) and (ii) as follows: \\
\begin{enumerate}
    \item[] \textbf{Case (i):\ } the effect of repulsive interactions between particles constituting the medium. It follows the negativity of the spring constant in Hooke's law $F=+kx$, where $x$ is a small deformation of the repulsive pairwise potential and $k$ is the strength of the locally downward parabolic potential $V_{\text{(i)}}=-\frac{k}{2}x^2$.\\
    \item[] \textbf{Case (ii):\ }  the spring constant is positive and the attractive interactions are described by a locally upward parabolic potential $V_{\text{(ii)}}=+\frac{k}{2}x^2$.\\
\end{enumerate}
In this work, we show that an inflation scenario occurs for case (i). 
We predict a rapid growth of the elastic stress–energy density during the inflationary phase, followed by its decay at the exit from inflation. During inflation, the scaling factor expands super-exponentially fast from $10^{-60}$ to $10^{-32}$ over a time $\sim 10^{-42}\text{s}$, see Fig. \ref{Fig2}. 
An alternative scenario would consider a transition between the two cases (i) and case (ii) during inflation. 
This hypothesis shows similar features as the theory of symmetry breaking between a \textit{false vacuum} and a true vacuum \cite{Starobinskii80,Guth81,Albrecht82,Linde82}. The best fit of luminosity curves using the parameters $K<0$ and $\e_0>0$ (see Fig. \ref{Fig1}) could be consistent with the hypothesis that the current situation corresponds to case (ii), and that a transition between cases (i) and (ii) is a feasible hypothesis.   
Theoretically,  we need to extend Hooke's law (see Eq. \eqref{Section:LambdaCDM:sigma}) to the non-linear regime to investigate such a transition. Searching for experimental evidence of cosmological inflation is a challenge for future research. Gravitational wave detectors offer new perspectives to explore the early universe. Unlike gravitational waves, the non-linear deformation waves that occur during inflation are longitudinal as the gauge $\partial_\mu \e^\mu_\nu = 0$ is no longer imposed and does not hold dynamically in the inflationary regime, see Eq. \eqref{Section:LambdaCDM:ModFriedmann2}. Hence, we expect longitudinal cosmic waves to be a signature of the inflation mechanism described in this article.   
 
As for the large order of magnitude of the bulk modulus, let us recall that classically an estimate shows that this value should be about $10^{20}$ times that of the steel, more specifically equal to $c^2 f^2/G \sim 4.5\times 10^{31} \text{N}\cdot \text{m}^{-2}$ for typical gravitational wave frequencies $f\sim 100 \text{Hz}$ \cite{McDonald18}. This value has been found using an analogy between the Einstein tensor with a stress tensor for shear deformation. In GR theory space-time is not a physical substance, and hence, the analogy with shear stress modulus that was drawn to derive the result is purely formal and not justified physically. 
However, we can compare our results with the values found using other approaches to the fabric of space-time. For example, the Young modulus was found to be of the order of $ c^7/(\hbar G^2) \sim 5\times 10^{113} \text{N}\cdot\text{m}^{-2}$, see \cite{Tenev18,David20}. This value can be understood as the \textit{Planck's Young modulus} and characterizes the stiffness of a quantum space-time. Consequently, this value does not correspond to an independent parameter in the cosmological model. On the contrary, in our present model, we introduce a new fundamental constant $K$ independent of any other constants. 

It is important to emphasize that our model does not reintroduce a classical ``ether'' in the sense of a preferred reference frame. The elastic medium is fully embedded within a covariant general relativistic framework. The deformation field $G^\mu$ is a covariant vector field, and the modified Einstein equations preserve diffeomorphism invariance. Moreover, the present-day deformation rate $\epsilon_0 \sim 10^{-133}$ implies that any observable effects, such as a fringe shift in a Michelson-Morley-type experiment, are far below current experimental sensitivity. Only cosmological-scale effects are expected to be detectable, assuming, of course, that the theory is correct.

Let us now comment on the differences between our model and the existing models proposed in the literature. The main difference between our approach and alternative gravity theories  \cite{Anisimov,QuintessencePeebles,QuintessenceCaldwell,QuintessenceBrax,QuintessenceReview} and dissipative cosmology \cite{Maartens,Fabris} is the idea of a universal coupling between the vector field $G_\mu$ and all kinds of matter-energy, see equations \eqref{Eq:LangrangianInteractionDefMatter}, \eqref{Eq:DefGeod:General}, and postulate (P3), that has not been proposed in the literature yet. The consequence of the equation \eqref{Eq:DefGeod:General} is that the modified (RFW) model has an additional equation in the system \eqref{Section:LambdaCDM:ModFriedmann3:v0}, which leads to an expression of the scaling factor that, to our knowledge, has not been proposed in this form in the existing literature (see equation \eqref{Section:LambdaCDM:ModFriedmann3}). The idea of violation of conservation energy has already been proposed in \cite{Perez17,Perez18}, however, their volume-preserving diffeomorphism condition does not allow them to explore such coupling between dark energy and matter-energy as we do in the present paper. This constraint is justified by the fact that their theory relies on the paradigm of metrical description of space-time and on the strong equivalence principle. As a result of this constraint, they find a change in the order of magnitude of the cosmological constant during inflation time that does not match the expected discrepancy. This highlights the limitation of describing inflation through a rigid/undeformable spacetime, while our model interprets inflation as the deformation of a cosmological medium (dark energy) with a very large bulk modulus.

Although our elastic-medium scenario introduces two macroscopic parameters $(K,\epsilon_0)$ (with the shear modulus $S$ set to zero for simplicity), it employs the same physical degree of freedom (i.e., the elastic deformation of a cosmological medium) to drive both primordial inflation and late-time acceleration. In that sense, it offers a unified dynamical origin for the two epochs, in contrast to the standard picture where an inflaton field and a separate cosmological constant (or quintessence sector) are invoked. 

{In conventional scalar-field inflation, the inflationary energy scale is set by
the height of the inflaton potential and may range from near-Planckian values down
to much lower scales, at the cost of increasing fine-tuning. In contrast, the
elastic-medium framework considered here does not rely on a scalar potential.
Instead, inflation is driven by the elastic stress--energy density
$\rho_\Lambda(t)c^2 = K\,\epsilon(t)$, where $K$ is the bulk modulus of the medium
and $\epsilon(t)$ its deformation rate.}

We acknowledge that a complete alternative to scalar field inflation must also account for the generation and evolution of primordial perturbations, particularly on superhorizon scales where physical observables become gauge-dependent. In particular, a successful model must explain the emergence of a nearly scale-invariant Harrison–Zel’dovich power spectrum \cite{Harrison1970, Zeldovich1972}, as predicted by quantum fluctuations in standard inflationary scenarios \cite{MukhanovChibisov1981, MukhanovReview1992}, and confirmed observationally with high precision by the Planck satellite, which finds $n_s = 0.9649 \pm 0.0042$ \cite{Planck2018}. In this work, we have focused on the background cosmological dynamics and the inflationary mechanism driven by elastic deformation. A perturbative analysis of our framework, one that carefully addresses gauge invariance \cite{Bardeen1980} and identifies observational signatures, is an essential next step. In a future companion paper, we aim to develop a consistent theory of linear perturbations in an elastic medium coupled to spacetime curvature. We anticipate that this approach may lead to new phenomenological predictions, such as residual longitudinal modes, which future CMB or gravitational wave experiments could probe.

{Beyond reproducing a nearly scale-invariant scalar spectrum, a crucial question
for any alternative to scalar-field inflation is its observational
distinguishability. In contrast with scalar-field and ekpyrotic models, the
elastic-medium framework naturally admits additional elastic degrees of freedom,
including longitudinal deformation modes associated with the elastic response of
the cosmological medium. These modes have no analogue in standard inflationary
scenarios and may lead to distinctive signatures in the tensor sector, such as
modified polarization content or departures from the usual scalar–tensor
consistency relations. While a full gauge-invariant perturbative treatment is
beyond the scope of the present work, identifying such characteristic features is
a primary motivation for the future development of the perturbation theory of the
model.
}\\

\textbf{Acknowledgments.} The author gratefully thanks the anonymous referees for insightful comments and constructive criticism that improved the clarity of this manuscript and highlighted directions for future work.

\bibliographystyle{IEEEtran}
\bibliography{APPB_Beau_v1}

\end{document}